\documentclass[11pt]{amsart}

\usepackage{hyperref}
\usepackage{amssymb,amsfonts,latexsym,amsmath,amsthm,wrapfig,graphicx}
\input xy
\xyoption{all}

\newtheorem{example}{Example}

\def\la{\label}
\def\be{\begin{equation}}
\def\beq{\begin{equation}}
\def\eeq{\end{equation}}
\def\ee{\end{equation}}
\def\bea{\begin{eqnarray}}
\def\eea{\end{eqnarray}}
\def\p{\partial}

\begin{document}

\title[Surface response analysis II]{Contributors of carbon dioxide in the atmosphere in Europe: the surface response analysis}

\author{Iuliana Teodorescu}
\author{Chris Tsokos} 
\address{Statistics Department, University of South Florida, Tampa Florida}

\begin{abstract}
This paper is a continuation of the statistical modeling of the nonlinear relationship between atmospheric CO$_2$ and attributable variables that can account for emissions, based on data from EU countries, in order to compare the relevant findings to those obtained in the case of US data, in \cite{model, article1}. The current study was initiated in \cite{article2}, leading to the optimal second-order model, based on three linear terms and five second-order terms. We conclude this study in the present work, by finding the canonical decomposition of the nonlinear model, and by computing the specific two-dimensional confidence regions that it leads to. We then use the model in order to quantify the net effect of various risk factors, and compare to the results obtained in the US case.  
\end{abstract}
\maketitle


\section{Introduction} \label{sec:intro} 

This article contains the second part of the statistical modeling of the nonlinear relationship between atmospheric CO$_2$ and various contributor variables that can account for emissions, based on data from EU countries. The first part \cite{article2} indicated the model-building procedure, including linear terms, quadratic terms, and mixed (interacting) terms, and produced rankings for the most significant attributable variables (or their interactions). 

In the current paper, we start from the second-order model developed in \cite{article2} and perform its surface reponse analysis, leading to canonical two-dimensional confidence regions, and to specific comparisons between canonical variables, much as it was done in \cite{article1}, in the case of US data. 

\begin{figure}[h!!!!!]
\begin{center}
 \includegraphics*[width=11cm]     {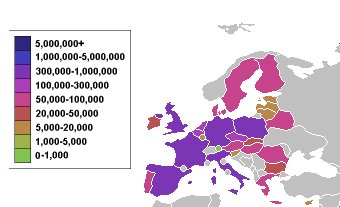}
\caption{EU CO$_2$ emissions, in thousands of metric tons.}
\label{map}
\end{center} 
\end{figure}

As indicated in the previous studies,  the response variable is the CO$_2$ in the atmosphere and is given in parts per million by volume (ppmv), obtained from yearly data  collected from 1959 to 2008\footnote{Year 1964 was ignored due to incomplete records.}. The CO$_2$ emission data for the EU countries listed below was obtained from Carbon Dioxide Information Analysis Center (CDIAC) during the same period: 
%
Austria,
Belgium,
Bulgaria,
Cyprus,
Czech Republic,
Denmark,
Finland,
France,
Germany,
Greece,
Hungary,
Ireland,
Italy,
Luxembourg,
Malta,
Netherlands,
Poland,
Portugal,
Romania,
Slovakia,
Spain,
Sweden,
United Kingdom 
(as of June 2013, there are 27 member states of the EU. Slovenia and the Baltic states were excluded from the present study since there was no individual data available for the period during which they were part of former Yugoslavia, and former Soviet Union, respectively. However, their contribution to the CO$_2$ emissions is relatively small, as it can be seen from Figure~\ref{map}\footnote{From \cite{pic}, reproduced by permission under the  Creative Commons Attribution-Share Alike 3.0 Unported license.}).

%
The goals of the surface response analysis for this problem are summarized below:

\begin{itemize}
\item starting from the second-order model derived in \cite{article2}, we perform a canonical decomposition of the quadratic part of the model. This will provide for us the relevant combinations of attributable variables (the canonical variables), and their respective effect (increasing, decreasing, or neutral) on the CO$_2$ emissions; 
\item depending on the different types of contributions at second-order level, we will classify and compute the various types of confidence regions, for pairs of canonical variables. The classification will produce confidence regions of elliptical and hyperbolic types, whose specific geometric parameters we will compute;  
\item finally, we use the results of the analysis to make recommendations for optimal management of various attributable variables, both from the point of emission reductions, and from that of ``cap-and-trade" policies, in order to optimize the energy and industry requirements of a state (or country) with respect to carbon emissions restrictions;
\item the study concludes with a descriptive comparison between the relevant attributable variables in the case of EU and US. We observe that there are significant differences between the most relevant variables (both at the level of single-factor and as interactions), and discuss possible consequences of interest for future policy development. 
\end{itemize}


\section{The model, parameters, and descriptive quantities}

We recall the final second-order model found in \cite{article2} to provide a good fit for the data and to have robust features for prediction and estimation:

\begin{eqnarray*}
[\widehat{CO}_2]^{-2.376}&  = & 0.00000123 + (710.85 Fl -30.64Ga -3.4501 Li )  \times 10^{-13} + \\
& + & (37.34Ga\cdot Bu + 1.35 Li \cdot Li -65.12 Bu \cdot Bu - \\
& - &  133.05 Li \cdot Fl -5.35 Li \cdot Bu)\times 10^{-18}.
\end{eqnarray*}

Throughout the paper, we will be using the the notation  $x_1 = $ Liquid Fuels (Li), $x_2 = $ Gas Fuels (Ga), $x_3 = $ Gas Flares (Fl), $x_4 = $ Bunker (Bu) for the relevant attributable variables. With respect to these variables, the model becomes

\be \la{first}
[\widehat{CO}_2]^{-2.376} = \beta_0 + \sum_{i=1}^4 \beta_i x_i + \sum_{i \le j = 1}^4 \beta_{ij}x_i x_j, 
\ee
with  the corresponding ranks determined by the stepwise  SAS procedure are given in Table 1, along with the coefficients in the final regression model.

In matrix notation  (where prime denotes transposition), \eqref{first} becomes

\be \la{model}
Y = \beta_0 + \beta' \cdot X + X' \cdot B \cdot X,
\ee
with the obvious identifications 

$$X' = (x_1, \ldots, x_4),  \,\, \beta' = (\beta_1, \ldots, \beta_4), \,\, B_{ij} = B_{ji} = \frac{1}{2}\beta_{ij}\,\, (i < j).$$

 \begin{center}
 \begin{table}[h!]
  \caption{Ranking by statistical relevance for attributable variables and interactions.}
 \begin{tabular}{| c || l | c | c |}
 \hline 
${\mbox{Rank}}$ & ${\mbox{Variable}} $ & $\beta \left [\times 10^{-18} \right ]$ & $F-$ Value  \\
 \hline 
\hline 
$  1 $  & Ga & $-30.635 \times 10^{5}$  & $197.22 $ \\
\hline 
$  2 $  & Ga:Bu & $37.3391$  & $50.25$   \\
\hline 
$  3 $  & Li:Li & $1.35565$  & $47.74 $   \\
\hline 
$  4 $  & Bu:Bu & $-65.115$  & $31.49$   \\
\hline 
$  5 $  & Fl & $710.848 \times 10^{5}$  & $26.98 $   \\
\hline 
$  6 $  & Li:Fl & $-133.05$  & $20.49 $  \\
\hline 
$  7 $  & Li:Bu & $-5.3501$  &   $19.07$ \\
 \hline 
$  8 $  & Li & $-3.4501 \times 10^{5}$  & $11.57$ \\
 \hline 
 \end{tabular}
 \end{table}
\end{center}

More precisely, the vector $\beta$ (up to an overall scale factor of $10^{-13}$), and the  symmetric matrix $B$ (up to an overall scale factor of $10^{-18}$) have the forms:

$$
\beta = 
\left [ 
\begin{array}{c}
-3.4501 \\  -30.635 \\ 710.848 \\ 0
\end{array}
\right ],
\, 
B = 
\left [ 
\begin{array}{cccccc}
2.7113 & 0 &  -133.05 & 0 \\
0 & 0 &  0 & 37.3391 \\
-133.05 & 0 &  0 & 0 \\
0 & 37.3391 & 0 & -130.23 \\
\end{array}
\right ]
$$

In order to perform the surface response analysis for this model, we must bring it to the simplest expression, by finding first its normal form and then its canonical decomposition.
Since these operations require inverting the matrix of second-order interactions, we need to perform a preliminary calculation in order to determine  its eigenvalues and corresponding orthonormal eigenvectors.

\subsection{Eingenvalue analysis of the second-order interactions matrix} \label{vect}

We recall that $\lambda_k, V_k$ ($k = 1, \ldots, 4$) are the eigenvalues and normalized eigenvectors of the  matrix $B$ if they solve the systems of linear equations:
$$
B \cdot V_k = \lambda_k V_k, \quad V'_k \cdot V_p = \delta_{kp}, 
$$
with $\delta_{ij}$ the Kronecker symbol, defined by $\delta_{ij} = 1$ if $i = j$ and $\delta_{ij} = 0$ otherwise. Then the matrix $B$ has the {\emph{principal-value decomposition}} (c.f. \cite[Appendix \S C]{west})
\be \la{pvd}
B = \sum_{k=1}^4 \lambda_k V_k V'_k.
\ee
For the matrix $B$ found above, upon computing numerically the eigenvalues (using the SAS RSREG procedure \cite{SAS} or Mathematica's Eigensystem procedure), 
we arrive at 
 \be \la{eval}
 \lambda_1 = -140.176, \,\, 
 \lambda_2  = 134.413, \,\, 
 \lambda_3 = -131.701, \,\, 
 \lambda_4 = 9.94612, \,\, 
\ee
up to the software numerical precision and the overall scale factor $10^{-18}$.  
     
The four orthogonal and normalized eigenvectors are found to be 
$$
V_1 = 
\left [ 
\begin{array}{c}
0\\ -0.257397 \\ 0 \\ 0.966306
\end{array}
\right ],
\,\, 
V_2 = 
\left [ 
\begin{array}{c}
0.7107\\ 0 \\ -0.703495 \\ 0 
\end{array}
\right ],
$$
$$
V_3 = 
\left [ 
\begin{array}{c}
-0.703495 \\ 0 \\ -0.7107 \\ 0 
\end{array}
\right ],
\,\, 
V_4 = 
\left [ 
\begin{array}{c}
0 \\ -0.966306 \\ 0 \\ -0.257397 
\end{array}
\right ].
$$

\subsection{Canonical analysis of the quadratic model}

Let $B^{-1}$ represent the inverse of the matrix $B$ (\cite[Appendix \S C]{west}) 
$$
B^{-1} = {\sum_{k=1}^4} \lambda_k^{-1} V_k V'_k, 
$$
and start from the model \eqref{model} 
$$
Y = \beta_0 + \beta' \cdot X + X' \cdot B \cdot X.
$$
In order to  bring this expression to its normal form, we begin by shifting the variable $X$ by a constant term 
$$
\widehat{X} = X + \frac{1}{2}B^{-1}\cdot \beta.
$$
Since $B$ is a non-singular matrix, we obtain the model 
$$
{Y} = \beta_0 + \beta' \cdot \widehat{X} - \frac{1}{4}\beta' \cdot B^{-1}\cdot \beta + \widehat{X}'\cdot B \cdot \widehat{X}
- \beta' \cdot B\cdot B^{-1} \widehat{X}, 
$$
where we have used the property $B^{-1} \cdot B  = \mathbb{I}$. Therefore, 
$$
{Y} = \beta_0  -\frac{1}{4}\beta' \cdot B^{-1}\cdot \beta +   \widehat{X}'\cdot B \cdot \widehat{X},
$$
so  we are now working with the normal quadratic form $\widehat{X}'\cdot B \cdot \widehat{X}$.  
Using again \eqref{pvd}, the quadratic form $\widehat{X}'\cdot B \cdot \widehat{X}$ becomes
$$
\widehat{X}' \left ( \sum_{k=1}^4 \lambda_k V_k V'_k \right ) \widehat{X} = \sum_{k=1}^4 \lambda_k  (\widehat{X}'V_k)(V'_k \widehat{X}) = 
\sum_{k=1}^4 \lambda_k  |V'_k\cdot \widehat{X}|^2 = \sum_{k=1}^4 \lambda_k z_k^2,
$$
where we have introduced the {\emph{canonical coordinates}} 
\be \la{z}
z_k := V'_k \cdot \widehat{X}, \quad k = 1, 2, 3, 4.
\ee

To conclude, we have the canonical form of the model

\be \la{canonical}
Y - Y_0 =  (-140.176 z_1^2  + 134.413 z_2^2 -131.701 z_3^2 + 9.94612  z_4^2 ) \times 10^{-18},
\ee
with $ z_k$ given in \eqref{z}.

To find the {\emph{stationary point}} of the model, defined generically  as the zero-gradient point, we must solve simultaneously for all $k = 1, \ldots, 4$:
$$
\frac{\p Y}{\p x_k} = 0 \Rightarrow \beta' + 2X' \cdot B = 0 \Rightarrow B\cdot X = -\frac{1}{2} \beta,
$$
which is equivalent to 
$$
B\cdot \widehat{X} = 0 \Rightarrow \widehat{X} = 0,
$$
because $B$ is non-degenerate. Together with \eqref{z}, this gives the stationary point as the origin 
of the $z$ coordinates, $z_1 = z_2 =  z_3= z_4 = 0$. In the original variables, the stationary point is found to be:
\be \la{sta}
X_s = -\frac{1}{2} B^{-1} \cdot \beta = 
\left [ 
\begin{array}{c}
-534271 \\ -286155 \\ -8294.32 \\ -82045.4
\end{array}
\right ],
\ee
up to an overall scale factor of $10^5$.

\subsection{Confidence region shapes and conic sections} 
We repeat here the discussion regarding confidence region types presented in \cite{article1}. 
In order to distinguish between various types of shapes the confidence regions may have, we now specialize to a pair of variables $(z_i, z_j) $ from the normal quadratic form written in canonical variables, and impose the inequality
$$
|Y - Y_0|  \le M, \quad M > 0,
$$
leading to
$$
\Big | \lambda_i z_i^2 + \lambda_j z_j^2 \Big | \le M, 
$$
which defines the confidence region centered at $(0, 0)$. We find the following cases, corresponding to classes of conic sections:

\begin{figure}
\begin{center}
 \includegraphics*[width=9.5cm]     {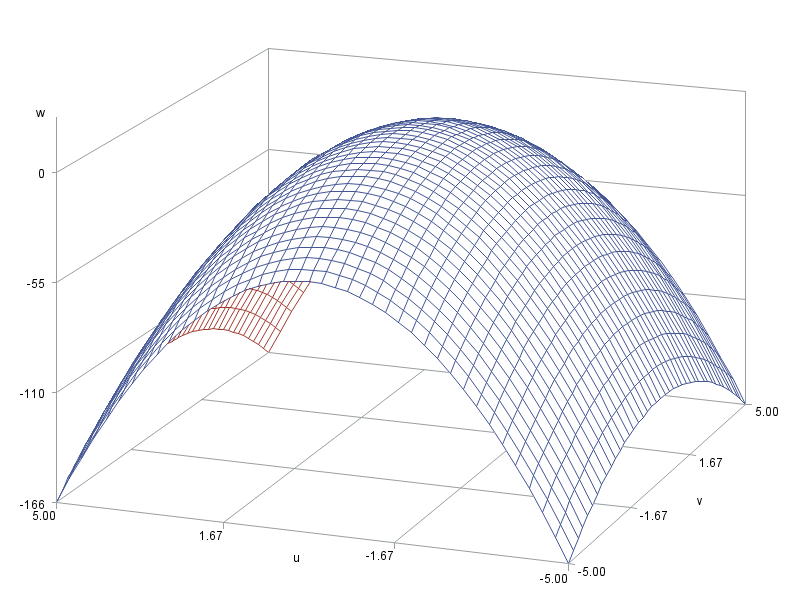}
 \includegraphics*[width=3cm]     {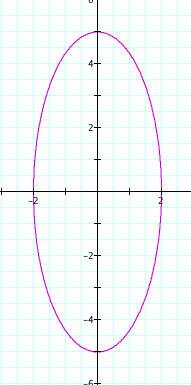}
\caption{Confidence regions for the elliptical case.}
\label{abs}
\end{center} 
\end{figure}

\subsubsection{Extremum point, elliptical region: all eigenvalues have the same sign}

If $\lambda_{i, j}$ are either all positive or all negative, the point $(0, 0)$ is a point of minimum or of maximum, respectively. The inequality becomes 
\be 
|\lambda_i| z_i^2 + |\lambda_j | z_j^2 \le M \Rightarrow \frac{z_i^2}{M/ |\lambda_i|} + \frac{z_j^2}{M/ |\lambda_j|} \le 1,
\ee
which defines the interior of an ellipse of semiaxes $\sqrt{M/ |\lambda_{i}|}, \sqrt{M / |\lambda_{j}|}$ (see Figure~\ref{abs}, right panel). 
The confidence region is given parametrically by:
\be \la{el}
z_i = \sqrt{\frac{M}{|\lambda_{i}|}} r \cos(\theta), \quad 
z_j = \sqrt{\frac{M}{|\lambda_{j}|}} r \sin(\theta), \quad 
0 \le r \le 1, \,\, 
\theta \in [0, 2\pi].
\ee
This is applicable for any pair of eigenvalues from $\{ \lambda_1, \lambda_3 \}$ or from $\{ \lambda_2, \lambda_4 \}$.

\begin{figure}[h!!!!!]
\begin{center}
 \includegraphics*[width=7.5cm]     {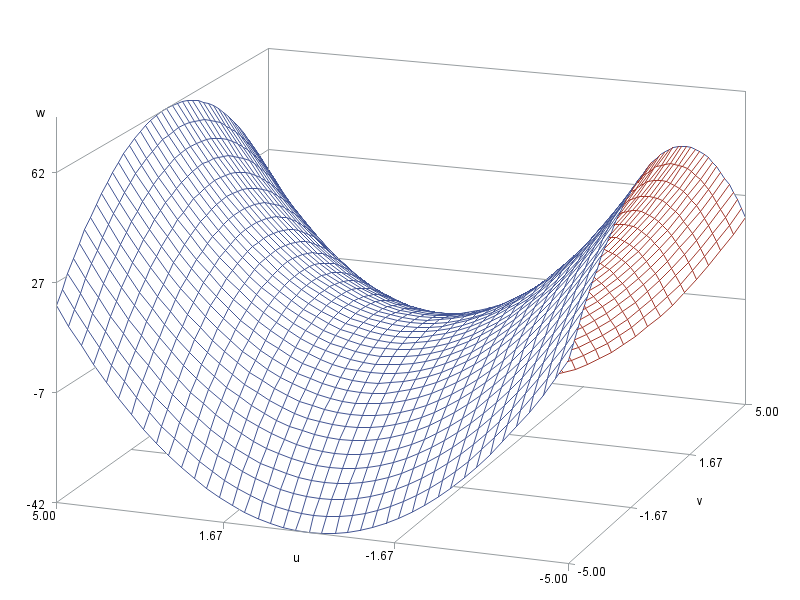}
 \includegraphics*[width=4.5cm]     {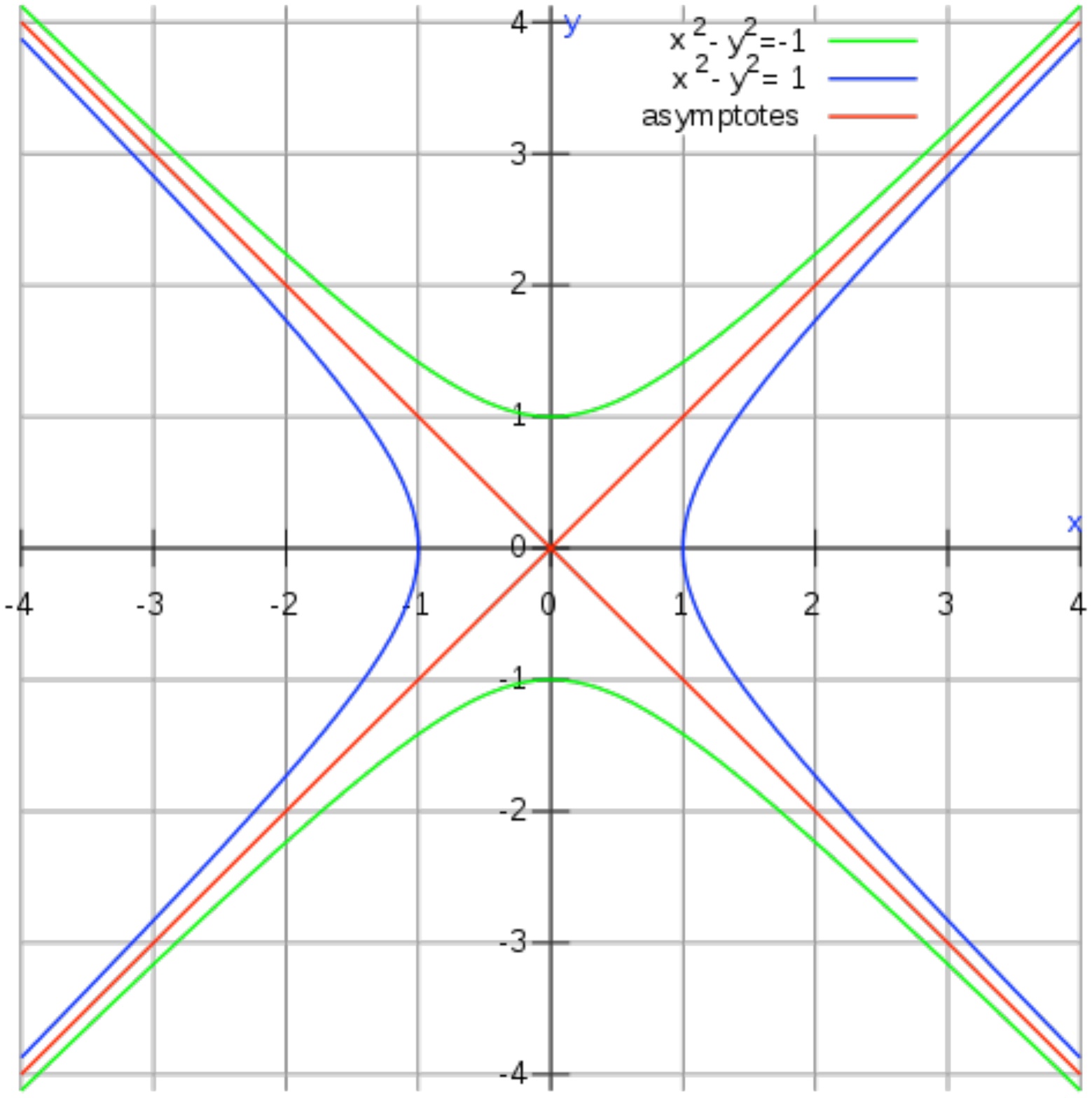}
\caption{Confidence regions for the hyperbolic case.}
\label{walker2}
\end{center} 
\end{figure}

\begin{example}
Determining specific numerical regions for CO$_2$ fluctuations at levels discussed by IPCC \cite{ipcc}.
\end{example}
In order to maintain consistency in comparing the models obtained for US \cite{article1} versus EU (this work), we compute the parameters of elliptical confidence regions for variables $z_1, z_3$ and $z_2, z_4$, corresponding to yearly CO$_2$ level fluctuations of 3\% (see the discussion in \cite[\S4.1]{article1} and supporting documentation in \cite{ipcc}). As shown in \cite[\S4.1]{article1}, this range of values corresponds to the order of magnitude $M \sim 10^{-8}$, so we arrive at the equations 
\be
0.7107 \widehat{X}_1- 0.703495 \widehat{X}_3 = 0, \quad -0.966306 \widehat{X}_2 - 
  0.257397 \widehat{X}_4 = 0, 
\ee
\be -0.257397 \widehat{X}_2 + 0.966306 \widehat{X}_4 = 
 8446.24\cdot r \cos(\theta),  
\ee
\be 
-0.703495 \widehat{X}_1 - 0.7107 \widehat{X}_3 = 8713.76 \cdot r\sin(\theta),
\ee 
which give the solution
\be
x_1 = 534271 - 6130.09\cdot r\sin(\theta), 
\ee
\be
x_2 = 286155 - 2174.03\cdot r\cos(\theta),
\ee
\be
x_3 = 8294.32 - 6192.87\cdot r \sin(\theta),
\ee
\be
x_4 = 82045.4+ 8161.64\cdot r \cos(\theta),
\ee 
where $ 0 \le r \le 1, \theta \in [0, 2\pi]$. It is important to note that this polar parametrization (in terms of the polar coordinates $r, \theta$) provides us with a confidence region more restrictive than just a product of maximal confidence intervals for individual variables $x_1, x_2, x_3, x_4$. The maximal confidence intervals would simply be
\be
|x_1 - 534271| \le  6130.09, \,\, |x _2 -2286155| \le 2174.03, 
\ee
\be |x_3 - 8294.32 | \le 6192.87, 
\,\, |x_4- 82045.4| \le 8161.64,
\ee
but the actual elliptical region will \emph{not} include the set of minimal values $x_1 = 528141, x_2 = 283981, x_3 = 2101.45, x_4 = 73883.8$, for instance.
\subsubsection
{Saddle-point, hyperbolic region: non-zero eigenvalues of different signs}

If, say, $\lambda_i > 0$ and $\lambda_j < 0$, then $(0, 0)$ is a saddle point, and  the inequality becomes 
$$
-M \le |\lambda_i| z_i^2 - |\lambda_j | z_j^2 \le M, 
$$
which defines the set of {\emph{orthogonal}} hyperbolas (see Figure~\ref{walker2})
\be \la{new}
\frac{z_i^2}{M/ |\lambda_i|} - \frac{z_j^2}{M/ |\lambda_j|} \le 1, \quad 
\frac{z_j^2}{M/ |\lambda_j|} - \frac{z_i^2}{M/ |\lambda_i|} \le 1.
\ee
The intersection of these conditions defines a region that looks like an elongated rectangle (elongated ``corners", the domain defined by the blue and green curves in Figure~\ref{walker2}) and can be approximated with a rectangular shape.
The confidence region is given parametrically by:

\be \la{hy}
z_i = \sqrt{\frac{M}{|\lambda_{i}|}} r \cosh(t), \quad 
z_j = \sqrt{\frac{M}{|\lambda_{j}|}} r \sinh(t), \quad 
-1 \le r \le 1, \,\, 
t \in \mathbb{R}.
\ee
This would give confidence regions for any choice  $\lambda_i \in \{ \lambda_1, \lambda_3 \}$ and $\lambda_j \in \{ \lambda_2, \lambda_4 \}$.

\begin{example}
As before, we compute specific confidence regions corresponding to the IPCC recommended values for yearly CO$_2$ fluctuations. 
\end{example}

Repeating the calculation performed in the previous example, for the case of hyperbolic confidence regions, we obtain (again, for $M \sim 10^{-8}$) the conditions
\be \la{q1}
x_1 = 534271 - 6130.08\cdot r\sinh(t), 
\ee
\be \la{q2}
x_2 = 286155 - 2174.03\cdot r\cosh(t),
\ee
\be \la{q3}
x_3 = 8294.32 - 6067.93\cdot r \sinh(t),
\ee
\be \la{q4}
x_4 = 82045.4+ 8161.64\cdot r \cosh(t),
\ee 
with $-1 \le r \le 1, t \in \mathbb{R}$. 

Notice that this does not provide an actual confidence region (the domain defined is unbounded), consistent with the geometric features shown in Figure~\ref{walker2}. 

However, we can extract from the conditions above specific linear relationships between the variables that can be used for comparison purposes. 
 Such linear relationships (which correspond to the asymptotic lines shown in Figure~\ref{walker2}, second panel) can be used to find equivalencies between variables $x_1, x_3$ and $x_2, x_4$. We perform this numerical analysis in Section \ref{num}, and indicate how to interpret the results.
%
%
%
%
%
%

%
%
%

\section{Conclusions and predictions based on nonlinear analysis} 

Throughout this subsection, we let the values of the attributable variables $X' = (x_1, x_2, x_3, x_4)$ be measured from the stationary point 
$X_s = -\frac{1}{2}B^{-1}\cdot \beta$   \eqref{sta}.

\subsection{Nonlinear analysis of contributing factors}

Starting from  \eqref{canonical}
$$
Y - Y_0 =  (-140.176 z_1^2  + 134.413 z_2^2 -131.701 z_3^2 + 9.94612  z_4^2 ) \times 10^{-18},
$$
and the power-law transformation 
$$
Y = ({\mbox{CO}}_2)^{-2.376},
$$
we first make the important remark that increasing/decreasing CO$_2$ is equivalent to decreasing/increasing $Y$. 

Next, using the defining relations for the linear combinations $z_k  = V'_k\cdot \widehat{X}$,  with $V_k$ given in \S \ref{vect}, we notice that the combinations $z_1, z_3$ contribute to increase the CO$_2$ emissions via interactions, while 
$z_2, z_4$ actually {\emph{decrease}} it. 
Given that (measured from the stationary point $X_s$),
$$
z_1 = -0.257397Ga + 0.966306Bu,
\quad
z_2 = 0.7107Li  -0.703495Fl, 
$$
$$
z_3 = -0.703495Li -0.7107Fl,
\quad
z_4 = -0.966306 Ga -0.257397Bu,
$$
we notice that $z_1$, which is mostly a combination of Gas Fuels and Bunker, has the most damaging effect.  Along with the fact that $x_1$ (Gas Fuels) ranks first among significant attributables in the second-order model, we can conclude that Gas-related sources seem to be the most significant factors responsible to the atmospheric CO$_2$ for the European countries studied here. 
%
%

\subsection{Relative importance of attributable variables} \la{num}

Finally, we can estimate the correct combinations between attributable variables $x_1 - x_4$ which would keep the CO$_2$ level constant, based on our model. It is particularly useful to observe that the variables $z_1, z_4$ are linear combinations only of attributables Ga, Bu, while $z_2, z_3$ are derived from the attributables Li, Fl. Therefore, it is natural to  impose the conditions
$$
\lambda_2 z_2^2 - |\lambda|_3 z_3^2 = 0, \quad -|\lambda_1| z_1^2 + \lambda_4 z_4^2 = 0,
$$
from which we obtain the hyperplane equations
$$
z_1 = \pm \sqrt{\left | \frac{\lambda_4}{\lambda_1} \right |}z_4, \quad 
z_2 = \pm \sqrt{\left | \frac{\lambda_3}{\lambda_2} \right |}z_3.
$$
These equations (using \S \ref{vect}) lead to the linear relationships between Ga-Bu and Li-Fl given below:
$$
-0.257397Ga + 0.966306Bu = \pm 0.266372913( -0.966306 Ga -0.257397Bu)
$$
$$
0.7107Li  -0.703495Fl = \pm 0.989860283 (-0.703495Li -0.7107Fl).
$$
From these equations it is possible to develop an equivalence between different attributables, and to use such identities in order to develop policy and accountability criteria.  The only acceptable solutions (selected by positivity of proportionality coefficients) yield:
\be
Ga = 1.74388 Bu, \quad Li = 98.1284 Fl.
\ee

In other words, under a ``CO$_2$ trade" policy developed under these guidelines, one unit of Gas fuel is equivalent to 1.74388 unit of Bunker, while one unit of Liquid Fuel can be replaced by 98.1284 units of Gas Flares. It is important to note 
that this ``conversion formula" corresponds to the condition $M = 0$, i.e. no variation in the CO$_2$ levels. For any other value of $M$, the formulas would provide different conversion values, as we show below. 

\subsection{Comparing the US and EU models}

Using the results obtained in \cite{article1}, it is possible develop a comparison between the US and EU quadratic models for attributable variables and interactions; in particular, it is possible to compare the relative relevance of the main single-factor variables and of the main interactions (see Table~\ref{comp}). 

 \begin{center}
 \begin{table}[h!]
  \caption{Comparison of statistical relevance for attributable variables and interactions, US vs. EU.}
  \label{comp}
 \begin{tabular}{| c || c | c |}
 \hline 
${\mbox{Rank}}$ & ${\mbox{Variable in US}} $  & ${\mbox{Variable in EU}} $  \\
 \hline 
\hline 
$  1 $  & Liquid  & Gas \\
\hline 
$  2 $  & Liquid:Cement & Gas:Bunker  \\
\hline 
$  3 $  & Cement:Bunker  & Liquid:Liquid \\
\hline 
$  4 $  & Bunker  & Bunker:Bunker \\
\hline 
$  5 $  & Cement   &  Gas Flares \\
\hline 
$  6 $  & Gas Flares   & Liquid:Gas Flares  \\
\hline 
$  7 $  & Gas  & Liquid:Bunker \\
 \hline 
$  8 $  & Gas:Gas Flares &  Liquid \\
 \hline 
 \end{tabular}
 \end{table}
\end{center}



In order to complete the comparison between the US and EU models, initiated in \cite{article1}, we evaluate conversion rates between attributable variables, corresponding to the same range of CO$_2$ level fluctuations as mandated by the IPCC ($M \sim 10^{-8}$ as shown in \cite[\S4.1]{article1}). As mentioned above, for a given value of $M \ne 0$, the conversion rates found earlier (for $M=0$) are not valid anymore. Instead, we start from the relations \eqref{q1}-\eqref{q4} (derived specifically for $M \sim 10^{-8}$), and arrive at the linear relations
established from these models: 
\be
\frac{x_1 - 534271}{6130.08} = \frac{x_3 - 8294.32 }{6067.93} = r \sinh(t),
\ee
\be
-\frac{x_2 - 286155}{ 2174.03} = \frac{x_4 - 82045.4}{8161.64} = r \cosh(t).
\ee

Therefore, we conclude that  under these conditions, one unit variation of $x_1$ (Liquid) corresponds to $6130.08/6067.93 \simeq 1.01$ units variations of $x_3$ (Gas Flares). Recall that in the US study \cite{article1} we concluded that 1000 units of Gas Flares can be equated to 2127 units of Cement;  the current study shows that in the case of EU, one unit of Liquid is equivalent to approximately 1.01 units of Gas Flares. However, a direct comparison of the various trading values cannot be derived, which is yet another indication that such studies must be performed regionally, and that application of uniform policies is not supported by the data.

\end{document}